\documentclass[a4paper,twoside]{article}
\newcommand{\affil}[1]{$^{\rm #1}$}
\usepackage[authoryear]{natbib}
\bibpunct{(}{)}{;}{a}{}{,}
\usepackage{graphicx}
\date{} 
\title{\large\bf\flushleft A photometric study of the newly discovered eclipsing cataclysmic variable 
SDSS J040714.78-064425.1}
\author{\parbox{\textwidth}{\flushleft
\vspace{-0.5cm}
{\it T.~Ak\affil{1}, A.~Retter\affil{2,3}, A.~Liu\affil{4}, and H.H.~Eseno\u{g}lu\affil{5}}\\
\vspace{0.4cm}
{\small \affil{1}\,Istanbul University, Faculty of Science, Dept. of Astronomy 
                   and Space Sciences, 34119, University, Istanbul, Turkey 
                   (tanselak@istanbul.edu.tr)}\\
{\small \affil{2}\,Pennsylvania State University, Department of Astronomy and 
                   Astrophysics, 525 Davey Lab., University Park, PA 16802-6305, 
                   USA (retter@astro.psu.edu)}\\
{\small \affil{3}\,University of Sydney, School of Physics, Sydney, NSW 2006, Australia}\\
{\small \affil{4}\,Norcape Observatory, PO Box 300, Exmouth, 6707, Australia 
                   (asliu@onaustralia.com.au)}\\
{\small \affil{5}\,Istanbul University Observatory Research and Application Center, 
                   34119 Beyaz$\i$t, Istanbul, Turkey}}}
\begin{document}
\twocolumn[
\begin{changemargin}{.8cm}{.5cm}
\begin{minipage}{.9\textwidth}
\vspace{-1cm}
\maketitle
%
%
\small{\bf Abstract: }We present the results obtained from unfiltered 
photometric CCD observations of the newly discovered cataclysmic variable 
SDSS J040714.78-064425.1 made during 7 nights in November 2003. 
We establish the dwarf nova nature of the object as it was in 
outburst during our observations. We also confirm the presence of 
deep eclipses with a period of $0.17017^{d}$$\pm$0.00003 in the optical 
light curve of the star. In addition, we found periods of $0.166^{d}$$\pm$0.001
and possibly also $5.3^{d}$$\pm$0.7 in the data. 
The $0.17017^{d}$ periodicity is consistent within the errors with the proposed 
orbital period of $0.165^{d}$ \citep{Szkodyetal2003} and $0.1700^{d}$ 
\citep{Monard2004}. Using the known relation between the orbital and superhump periods, 
we interpret the $0.166^{d}$ and $5.3^{d}$ periods as the negative superhump and the nodal 
precession period respectively. SDSS J040714.78-064425.1 is then classified 
as a negative superhump system with one of the largest orbital periods.


\medskip
\medskip
\end{minipage}
\end{changemargin}
]
\small

\section{Introduction}
SDSS J040714.78-064425.1 (hereafter abbreviated to SDSS J0407-0644, 
$\alpha_{2000.0}$=$04^h$$07^m$$14.78^s$, 
$\delta_{2000.0}$=-$06^{\circ}$ $44^{'}$$25.1^{"}$,
\citealt{Downesetal1997}) is a cataclysmic variable 
star discovered from SDSS (Sloan Digital Sky Survey,  
\citealt{Yorketal2000}) by \citet{Szkodyetal2003}. 
They also obtained V filter photometry 
and time-resolved spectroscopy of the system while its 
out-of-eclipse magnitude was about 17.2 mag. According to 
Szkody et al., SDSS J0407-0644 is a deeply eclipsing cataclysmic 
variable with an eclipse depth of about 2 mag. Their 
photometric data revealed an orbital period of about 3.96 hr and 
a strong orbital hump modulation interpreted by a prominent hot spot. 
However, their observations only cover 9.42 hrs, i.e. 2 eclipses, 
so the error in this orbital period determination is quite large.  
Spectra of SDSS J0407-0644 show the double-peaked Balmer line 
profile during the out-of-eclipse phases, typical of
accretion discs. The blue peak is stronger during the hump 
at phase 0.9, and the Balmer emission is unchanged during the 
eclipse itself (phase 0.0). 

\citet{Monard2004} reported that the star showed 
several normal dwarf nova outbursts since July 2003. 
His observations also indicated two possible superoutbursts 
around 19 July and 17 October, 2003 suggesting that the system 
may be classified as an SU UMa type dwarf nova candidate. 
\citet{Henden2004} reported outbursts as well.

The possibility that SDSS J0407-0644 is an SU UMa system with an 
orbital period above the period gap inspired us to observe 
this system in an outburst and to look for superhumps.

In this paper, we present extensive photometric 
observations of SDSS J0407-0644, which suggest a refined orbital 
period, a negative superhump period and possibly a precession 
period as well.

\section{Observations}
Photometric observations of SDSS J0407-0644 were made by one of us, Liu, 
with a 0.3-m f/6.3 telescope coupled to an ST7 NABG CCD camera.
The telescope is located in Exmouth, Western Australia, and no filter 
was used. Exposure times were 60 sec every 120 sec. 
SDSS J0407-0644 was observed in the nights of 5, 6, 7, 8, 9, 11 and 12 
November, 2003. The observational log is given in Table 1. 
The observations span 7 nights (48 hours in total). 

We estimated differential magnitudes with respect to GSC4731-0533 
(V=$14.30^{m}$, the comparison star, 2.1' N of the variable), using 
another star (1.57' SE of SDSS J0407-0644) in the field, which is not listed 
in the Guide Star Catalog, as the check star. The unfiltered magnitude 
of the check star ($16.60^{m}$) was derived from the SBIG CCDOPS software 
that came with the camera.

\begin{table}[h]
\begin{center}
\caption[]{Journal of the photometric observations made on November 2003. 
           N and $m_{out}$ denote the number of observations and the mean 
           out-of-eclipse magnitudes, respectively.}
\tiny
\begin{tabular}{lccccc}
\hline
Day        & HJD Start     & Duration &   N  &  mid-eclipse  & $m_{out}$ \\
           & (HJD-2452900) & (hours)  &      & (HJD-2452900) &  (mag)   \\
\hline
 05 & 49.09812      &  6.3     & 168  &   49.15016    &   15.48  \\
    &               &          &      &   49.32062    &           \\
 06 & 50.08346      &  6.7     & 185  &   50.17176    &   15.62  \\
    &               &          &      &   50.34135    &           \\
 07 & 51.05323      &  7.5     & 207  &   51.19289    &   15.70  \\
    &               &          &      &   51.36300    &           \\
 08 & 52.08026      &  6.8     & 184  &   52.21371    &   15.92  \\
 09 & 53.07978      &  6.8     & 170  &   53.23440    &   16.06  \\
 11 & 55.07301      &  6.8     & 180  &   55.10714    &   16.75  \\
    &               &          &      &   55.27290    &           \\
 12 & 56.06455      &  7.1     & 193  &   56.12916    &   17.11  \\
    &               &          &      &   56.29833    &           \\
\hline
\end{tabular}
\end{center}
\end{table}

Differential magnitudes were calculated using aperture photometry. 
The mean GSC magnitude of the comparison star was added to the 
differential magnitudes to give a rough estimate of the visual 
magnitude. The light curve of SDSS J0407-0644 obtained during the 
observations is shown in Figure 1. A part of the light curve of SDSS J0407-0644 
observed on November 05, 2003 is presented in Figure 2. The observational 
errors were estimated from the deviations of the K-C magnitudes from the
nightly means and are typically about 0.03 mag.

As can be seen in Figure 1, the mean out-of-eclipse unfiltered magnitude of 
SDSS J0407-0644 decreased from 15.48 to 17.11 mag during the observing run. 
\citet{Monard2004} found the system in possible superoutbursts 
on July 19 and October 17, 2003, while its mean out-of-eclipse magnitudes 
was about 15.15 and 14.98 mag, respectively. However, the out-of-eclipse 
magnitude of SDSS J0407-0644 was about 17.7 mag in the observations of 
\citet{Szkodyetal2003} performed in November 14, 2002 (see their Figure 10). 
These values show that we observed the system declining from an outburst, 
while Szkody et al. studied SDSS J0407-0644 in quiescence.

\begin{figure}[h]
\begin{center}
 \includegraphics[scale=0.27, angle=0]{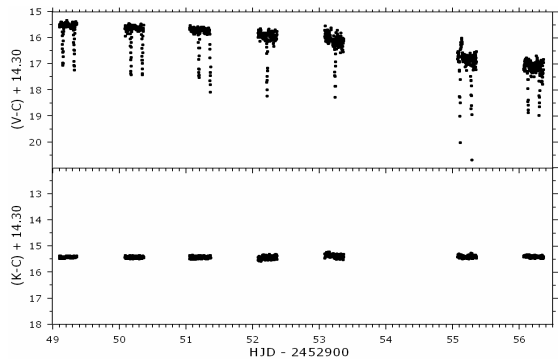}
\caption{\small The light curve of SDSS J0407-0644 during the observing 
            run. "V", "C" and "K" represent unfiltered magnitude 
            of the variable, comparison and check stars, respectively. 
            The mean GSC magnitude of the comparison star was added 
            to the differential magnitudes to give a rough estimate 
            of the visual magnitude of SDSS J0407-0644 and the check star.}\end{center}
\end{figure}


\begin{figure}[h]
\begin{center}
\includegraphics[scale=0.27, angle=0]{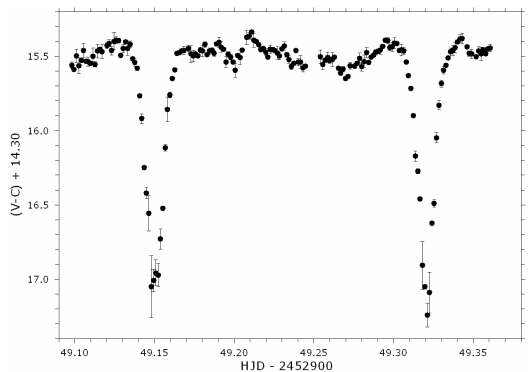}
\caption{\small A part of the light curve of SDSS J0407-0644 observed on 
            November 05, 2003.}
\end{center}
\end{figure}

\section{Analysis}

\subsection{The periodogram analysis}

The period analysis was performed using the Data 
Compensated Discrete Fourier Transform 
(DCDFT, \citealt{FerrazMello1981}, \citealt{Foster1995}), 
including the CLEAN algoritm \citep{Roberstsetal1987}. The DCDFT 
method is based on a least-square regression on two 
trial functions, sin(ft) and cos(ft), and a constant. 
Here f denotes the frequency. In the period analysis, 
we assume that the frequency, say $f_{1}$, that corresponds to the 
highest peak in the power spectrum is real and subtract 
it from the data. Then, we find the highest 
peak, say $f_{2}$, in the power spectrum of the residuals, 
subtract $f_{1}$ and $f_{2}$ simultaneously from the raw data and 
calculate a new power spectrum etc. until the strongest residual peak is 
below a given cutoff level. To select the peaks, we followed 
a conservative approach which is similar to the method described 
by \citet{Bregeretal1993} who gave a good criterion for the 
significance of a peak in the power spectrum. 
In Breger's method, the peaks in the power spectrum which 
are higher than the signal to noise ratio, S/N, of 4.0 for 
the amplitude are indicators of real signals.
In order to assign a confidence level to the power 
spectra, we calculated the standard error ($\sigma$) of the 
power values between the frequencies for which no strong peaks 
appear. We assumed 4$\sigma$ to be the confidence level for 
the power and considered only those peaks of the power 
spectrum whose power was above this level. We then applied 
the CLEAN algorithm to remove the false peaks until the 
strongest residual peak is below the calculated confidence 
level. Note that we also searched for periodic brightness 
modulations by Period98 \citep{Sperl1998}, which 
is based on a least-square regression on a trial function, 
sin(ft), along with a zero point. We found very similar power 
spectra from both techniques. We calculated the error in a frequency 
from the half width at the half maximum of the peak which 
is a good rough estimator of the uncertainty in a frequency.

\begin{figure}[h]
\begin{center}
\includegraphics[scale=0.26, angle=0]{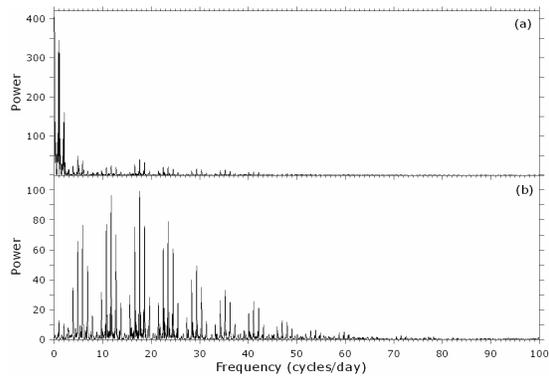}
\caption{\small Power spectra of SDSS J0407-0644.  
            $\bf (a)$ The power spectrum of the raw light curve. 
            $\bf (b)$ The power spectrum after the mean magnitudes 
                      of the single nights were subtracted from the 
                      observations.}
\end{center}
\end{figure}

\subsubsection{The raw data}

As can be seen in Figure 1, the mean out-of-eclipse magnitude of 
SDSS J0407-0644 was changed during the observing run.
This magnitude change creates a strong low frequency signal at 
the frequency 0.046 $\pm$ 0.002 c/d ($21.8^{d}$ $\pm$ 1.1) in the 
power spectrum. Thus, the power spectrum of the light curve of 
SDSS J0407-0644 is dominated by the aliases of this low frequency signal 
and the harmonics of the orbital frequency near 6 c/d as demonstrated 
in Figure 3. Since the observations were done from a single site, the 
1 c/d alias is also very strong, as expected.

\begin{figure}[h]
\begin{center}
\includegraphics[scale=0.27, angle=0]{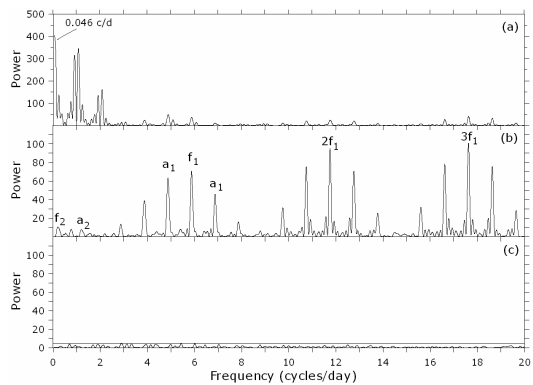}
\caption{\small Power spectra of SDSS J0407-0644 zoomed into the 0-20 c/d range of
	      frequencies. '$a_{i}$' (i=1-2) represent $1d^{-1}$ aliases 
            of '$f_{i}$'. 
            $\bf (a)$ The power spectrum of the raw light curve of 
                      SDSS J0407-0644. 
            $\bf (b)$ The power spectrum after fitting and 
                      subtracting the very low frequency signal 
                      of 0.046 c/d that originates from the general 
                      trend of the data. 
            $\bf (c)$ The cleanest power spectrum after fitting 
                      and subtracting $f_{1}$=5.868 (with its harmonics) and 
                      $f_{2}$=0.19 c/d. The confidence level of 5.08 
                      (4$\sigma$) for the power is shown with a horizontal line.}
\end{center}
\end{figure}

After fitting and subtracting the very low frequency signal 
of 0.046 c/d that originates from the general trend of the 
light curve, the orbital frequency and its harmonics at the 
frequencies 11.753 $\pm$ 0.009 and 17.628 $\pm$ 0.009 c/d
dominate the power spectrum of the residuals as shown in Figure 4b. 
The peak that corresponds to the orbital modulation is found at the 
frequency $f_{1}$=5.868 $\pm$ 0.009 c/d ($0.1704^{d}$ $\pm$ 0.0003).
In order to find a confidence level for the power, we calculated 
the standard error ($\sigma$) of the power level to be 1.27 between 
8-20 c/d after subtracting the harmonics of $f_{1}$.
By considering this standard error as the noise level, we 
calculated the confidence level to be 4$\sigma$=5.08 for the power, 
as described above. To search for additional signals, $f_{1}$ 
and its harmonics were also subtracted from the data. In the 
power spectrum of the residuals, the 
strongest peak corresponds to the frequency of $f_{2}$=0.19 $\pm$ 0.02 c/d 
($5.3^{d}$ $\pm$ 0.7). The amplitude of this signal is 0.16 $\pm$ 0.03 mag.
After removing $f_{2}$ from the data, we conclude 
that there is no additional significant signal in the light curve, since the power 
level in the residual power spectrum (Figure 4c) is lower than the 
confidence level. However, the presence of the $5.3^{d}$ is very 
questionable due to following reasons : (1) Since the data cover about 
$7.3^{d}$, we only have less than 1.5 cycles of the proposed $5.3^{d}$ 
period, (2) We subtracted the general trend of the light curve which 
may affect low frequencies. Thus, this low frequency signal can be 
an artifact of the removal of the 0.046 c/d signal, (3) For a decaying 
system one may expect low frequency signals.

\subsubsection{The frequencies near 4-8 c/d}

In order to test the significance of the peak found at 5.868 c/d and 
to search for additional signals near this frequency, we calculated the 
mean magnitudes of each night and subtracted them from the observations 
of SDSS J0407-0644.

\begin{figure}[h]
\begin{center}
\includegraphics[scale=0.27, angle=0]{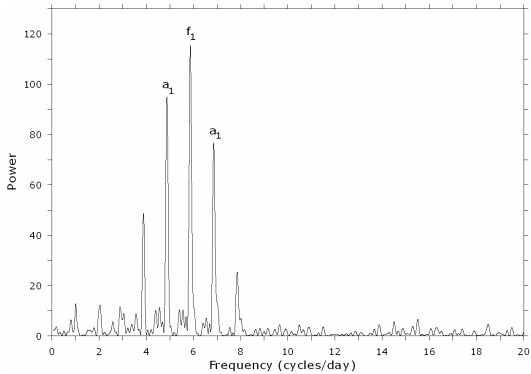}
\caption{\small The power spectrum of the light curve of SDSS J0407-0644 
            zoomed into the 0-20 c/d range after the mean out-of-eclipse 
            magnitudes of each night were subtracted from the observations. 
            The $\geq$2 harmonics of $f_{1}$ were removed from the power 
            spectrum for clarity. '$a_{1}$' represent $1d^{-1}$ aliases of 
            '$f_{1}$'.}
\end{center}
\end{figure}

\begin{figure}[h]
\begin{center}
\includegraphics[scale=0.27, angle=0]{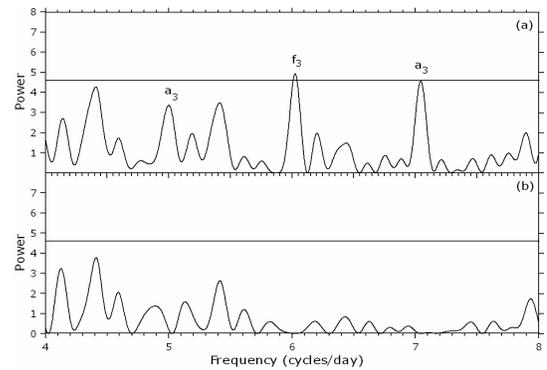}
\caption{\small The power spectra of the light curve of SDSS J0407-0644 zoomed 
            into the interval 4-8 c/d. The confidence level of 4.60 (4$\sigma$)
		is shown with horizontal lines. '$a_{3}$' represent $1d^{-1}$ 
            aliases of '$f_{3}$'. 
            $\bf (a)$ The power spectrum after $f_{1}$ and the 1 c/d alias 
                      are removed. 
            $\bf (b)$ The power spectrum after fitting and subtracting 
                      $f_{3}$=6.03 c/d as well.}
\end{center}
\end{figure}

The power spectrum of this de-trended data is dominated by the harmonics 
of the orbital frequency and the 1 c/d aliases which originate from the 
observational gap. The harmonics in the power spectrum are the result of 
the non-sinusoidal shape of the eclipses. The power spectrum is shown in Figure 5.
In this power spectrum, the peak that corresponds to the orbital modulation is found at 
the frequency 5.869 $\pm$ 0.009 c/d ($0.1704^{d}$ $\pm$ 0.0003) which is 
consistent with $f_{1}$ found above. By considering the standard error 
$\sigma$=1.15 of the power between 8-20 c/d as the noise level after 
subtracting the harmonics of $f_{1}$, we calculated the confidence level 
to be 4$\sigma$=4.60 for the power.

To search for additional signals, $f_{1}$, its harmonics and the 1 c/d aliases 
were subtracted from the data. The strongest peak in the power 
spectrum of the residuals corresponds to the frequency of 
$f_{3}$=6.03 $\pm$ 0.05 c/d ($0.166^{d}$ $\pm$ 0.001) which is demonstrated 
in Figure 6a. The amplitude of this signal is 0.04 $\pm$ 0.02 mag.
It can be seen in Figure 6a that the power level of this 
very weak peak is just above the confidence level. The light curve 
folded on the $0.1704^{d}$ and $0.166^{d}$ periods is shown in 
Figure 8a and b, respectively.

\begin{figure}[h]
\begin{center}
\includegraphics[scale=0.27, angle=0]{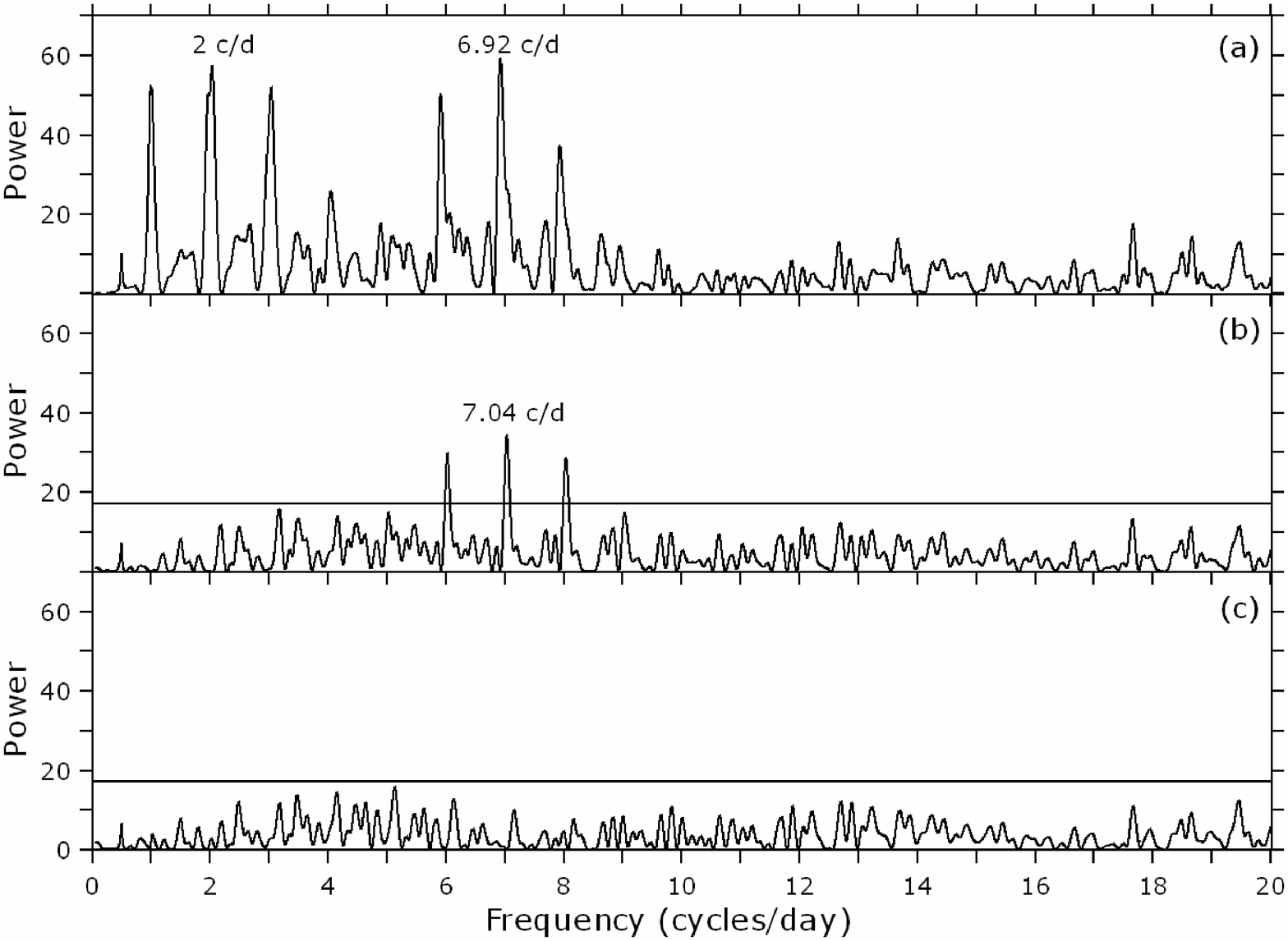}
\caption{\small The power spectra of the light curve of SDSS J0407-0644 
            zoomed into the 0-20 c/d range of frequencies after 
            removing the eclipses by hand. The confidence level 
            of 17.16 (4$\sigma$) is shown with horizontal lines. 
            $\bf (a)$ The power spectrum of the data. 
            $\bf (b)$ The power spectrum of the residuals after fitting 
                      and subtracting the peaks that correspond to the $1d^{-1}$ 
                      aliases of the orbital frequency (6.92 c/d) and the 
                      observational gaps (2 c/d).
            $\bf (c)$ The power spectrum of the residuals after fitting and 
                      subtracting the peak at 7.04 c/d as well.}
\end{center}
\end{figure}

\begin{figure}[h]
\begin{center}
\includegraphics[scale=0.27, angle=0]{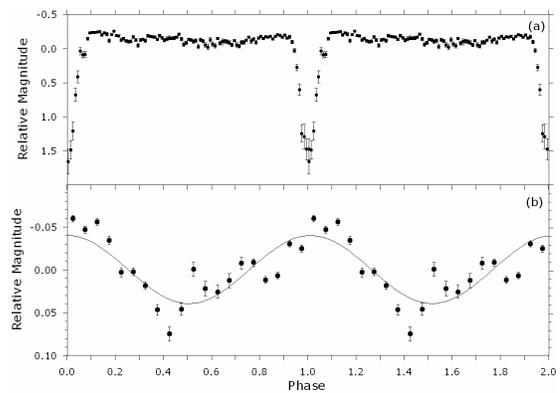}
\caption{\small The folded light curves. The data used here are 
            the observations after the nightly means were subtracted 
            (see Section 3.1.2). Error bars represent one standard 
            deviation of the mean values. 
            $\bf (a)$ The light curve folded on the $0.1704^{d}$ 
                      period and binned into 100 bins. 
            $\bf (b)$ The light curve folded on the 
                      $0.166^{d}$ period and binned into 20 bins 
                      after after the eclipses were removed by hand. 
                      The solid line shows the sine fit.}
\end{center}
\end{figure}

\subsubsection{The out of eclipse data}

Since the significance of the signal with the frequency 
$f_{3}$=6.03 c/d is questionable due to its low amplitude 
in the power spectra of the light curve, we removed the 
eclipses by hand in the de-trended data in order to test the
presence of this signal. 

Power spectra of the out-of-eclipse data are displayed in Figure 7. 
In the power spectrum, we find the strongest peak at the frequency 
6.92 $\pm$ 0.02 c/d which is near the $1d^{-1}$ 
alias of the orbital frequency 5.869 c/d. The $1d^{-1}$ alias 
of the frequency that corresponds to the observational gaps is 
also very prominent. Finding the strongest peaks at the frequencies 
that correspond to the $1d^{-1}$ aliases of the signals present 
in the data is not a surprise since we changed the window 
function of the data by removing the eclipses. 
After subtracting the $1d^{-1}$ aliases of the orbital frequency 
and the observational gaps, we find the strongest peak in the power 
spectrum of the residuals at the frequency 7.04 $\pm$ 0.02 c/d which 
is consistent with the $1d^{-1}$ alias of the candidate signal at 
the frequency $f_{3}$=6.03 c/d. Thus, we conclude that the 
$f_{3}$=6.03 c/d signal is indeed present 
in the light curve. In Figure 7b, the standard error $\sigma$ of 
the power in the frequency interval 0-20 c/d is 4.29. 
By considering $\sigma$ as the noise level, we calculated the 
confidence level to be 4$\sigma$=17.16 for the power.

\subsection{The ephemeris and eclipse \\ depths}

The observations cover 12 usable eclipses. The corresponding 
mid-eclipse timings were determined by fitting a Gaussian to the 
data near the eclipses and are given in Table 1. We applied a 
linear least-squares fit to the minima times and found the 
following orbital ephemeris:

   \begin{equation}
      T_{min}(HJD) = 2452949.1504(6) + 0.17017(3) E      
   \end{equation}

Since the error in this value is smaller than the result from 
the periodogram analysis where the error was defined as the 
half width of the peaks at half maximum, we adopt this value 
for the orbital period.

We also measured the eclipse depths by fitting a Gaussian to 
the eclipse profiles after we subtracted the mean magnitudes 
of each night from the observations. The eclipse depths given 
in \citet{Szkodyetal2003} were about 1.5 mag. However, the mean 
eclipse depth is 1.80 $\pm$ 0.24 mag in our observations and 
they range between 1.5 and 2.2 mag.

\section{Discussion}
 
The $0.17017^{d}$ period that we found in the data is consistent 
with the orbital period ($0.165^{d}$) suggested by \citet{Szkodyetal2003} 
who did not give an error estimation for the orbital period. Since our 
observing run is much longer than the study of Szkody et al. and since 
they used only two eclipses to determine the periodocity, while we 
used 12, our value is more precise. It should be noted that \citet{Monard2004} 
reported an orbital period of $0.1700^{d}$ $\pm$ 0.0002 from his 
observations that covered 4 eclipses obtained in Bronberg 
Observatory/CBA Pretoria. Thus, the $0.17017^{d}$ period found 
from the photometric observations of SDSS J0407-0644 in this study is 
naturally explained as the orbital period of the system. In short, 
our data confirm the presence of eclipses in the light curve of 
SDSS J0407-0644 as deep as 2 mag and this implies that the orbital 
period is $0.17017^{d}$.

The 6.03 c/d frequency corresponds to the period $0.166^{d}$ $\pm$ 0.001 
which is $\sim$3$\%$ shorter than the suggested orbital period of 
$0.17017^{d}$. Negative superhump periods are a few percent shorter 
than the orbital periods \citep{Pattersonetal1997,Patterson1999,Patterson2001,RetterandNaylor2000}. 
The $0.166^{d}$ period 
fits this trend and thus can be understood as the negative superhump 
period in SDSS J0407-0644. Negative superhumps are explained as the beat 
between the orbital period and the nodal precession of the disc. In 
this model, the relation between the orbital period $P_{orb}$, the 
negative superhump period $P^{-}_{sh}$ and the precession period of the 
disc $P_{pr}$ is given by $1/P_{pr}$=$1/P^{-}_{sh}$$-$$1/P_{orb}$, 
or $f_{pr}$=$f^{-}_{sh}$$-$$f_{orb}$. By choosing $f_{orb}$= 5.869 c/d 
and $f^{-}_{sh}$= 6.03 c/d from the power spectrum of SDSS J0407-0644 
(Section 3), we calculate a precession frequency of 
$f_{pr}$ =0.16 $\pm$ 0.04 c/d. Thus, we expect a signal with the frequency 
of $f_{pr}$ $\approx$ 0.16 c/d in the light curve of SDSS J0407-0644. This 
precession frequency is in agreement with the signal found from the 
light curve at $f_{2}$=0.19 $\pm$ 0.02 c/d (Section 3). 
However, the presence of the $5.3^{d}$ 
period is very questionable (see Section 3.1.1). 

From our results SDSS J0407-0644 can be classified as a negative superhump
system with one of the largest orbital periods. Only TV Col with an
orbital period of 5.5 h \citep{Retteretal2003} and 
AT Cnc at 4.8 h \citep{Nogamietal1999,Kozhevnikov2004} have longer orbital 
periods and show superhumps.

Positive superhumps are also observed in cataclys- mic variables. 
They are explained as the beat between the binary motion and the 
precession of the disc in the apsidal plane 
\citep{Patterson1999}. The positive superhump periods are 
a few percent longer than the orbital periods. Patterson found that 
the period deficits in the negative superhump systems are about half 
the period excesses in the positive superhump systems: 
$\epsilon_{-}$$\approx$$-$0.5$\epsilon_{+}$, where 
$\epsilon$=($P_{sh}$$-$$P_{orb}$)/$P_{orb}$. 
\citet{Retteretal2002} further suggested that 
the ratio $\epsilon_{-}$/$\epsilon_{+}$, which is expected 
to be 0.5 according to \citet{Patterson1999} is correlated with 
the orbital period. We found a negative superhump period deficit 
of $-0.026$ for SDSS J0407-0644 (Section 3). From the relations between the 
negative superhump deficit and the positive superhump excess 
mentioned above, we expect a positive superhump excess of about 
$\epsilon_{+}$$\approx$$+0.04-0.05$ for SDSS J0407-0644. 
This yields a positive superhump frequency around $5.58-5.63$ c/d. 
As can be seen in Figure 6 there is a very weak peak at the 
frequency 5.41 c/d. However, it is below the confidence level. Thus, 
we conclude that the system did not show a positive superhump or 
that its amplitude was below the detection limit in our observations.

The mean eclipse depth of the system in our data is 1.80 $\pm$ 0.24 mag. 
\citet{Szkodyetal2003} found an orbital period of 3.96 hr and 
a strong orbital hump modulation in the light curve of SDSS J0407-0644 
that included only two eclipses (see their Figure 10). However, 
the amplitude of orbital hump in our observations was about 
10 times smaller than that observed by Szkody et al. 
We observed SDSS J0407-0644 in an outburst while Szkody et al. studied 
it in quiescence (Section 2). Thus, the difference between the 
amplitudes of the out-of-eclipse humps obtained in this study 
and in Szkody et al. may be due to the difference in the brightness 
state of the system.

These are the first determinations of the superhump and 
the possible precession periods in the light curve of 
SDSS J0407-0644. Further observations are needed to confirm our results.

\section*{Acknowledgments}
We thank the anonymous referee for a thorough
report and useful comments that helped improving an 
early version of the paper. This work was partially 
supported by a postdoctoral fellowship from Penn State 
University and by the Research Fund of the University 
of Istanbul, Project Numbers: BYP-409/26042004 and 
BYP-239/0606082003.


\end{document}